# An ensemble approach for feature selection of Cyber Attack Dataset

Shailendra Singh
Department of Information Technology
Rajiv Gandhi Technological University
Bhopal, India
e-mail:shailendrasingh@rgtu.net

Sanjay Silakari
Department of Computer Science & Engineering
Rajiv Gandhi Technological University
Bhopal, India
e-mail:ssilakari@rgtu.net

*Abstract*— **Feature selection is an indispensable pre-processing step when mining huge datasets that can significantly improve the overall system performance. Therefore in this paper we focus on a hybrid approach of feature selection. This method falls into two phases. The filter phase select the features with highest information gain and guides the initialization of search process for wrapper phase whose output the final feature subset. The final feature subsets are passed through the K-nearest neighbor classifier for classification of attacks. The effectiveness of this algorithm is demonstrated on DARPA KDDCUP99 cyber attack dataset**.

I. INTRODUCTION

Feature selection aims to choose an optimal subset of features that are necessary and sufficient to describe the target concept. It has proven in both theory and practice effective in enhancing learning efficiency, increasing [1][2] predictive accuracy and reducing complexity of learned results. Optimal feature selection requires an exponentially large search space, where N is the number of features [3]. So it may be too costly and impractical. Many feature selection methods have been proposed in recent years. The survey paper [4] gives the complete scenario of different approaches used in cyber attack detection systems. They can fall into two approaches: filter and wrapper [5]. The difference between the filter model and wrapper model is whether feature selection relies on any learning algorithm. The filter model is independent of any learning algorithm, and its advantages lies in better generality and low computational cost [6]. The wrapper model relies on some learning algorithm, and it can expect high classification performance, but it is computationally expensive especially when dealing with large scale data sets [7] like KDDCUP99.

This paper combines the two models to make use of their advantages. We adopt a two-phase feature selection method. The filter phase selects features and uses the feature estimation as the heuristic information to guide wrapper algorithm. We adopt information gain [8] uncertainty to get feature estimation. The second phase is a data mining algorithm which is used to estimate the accuracy of cyber attack detection. We use K-nearest neighbor based wrapper selector. The feature estimation obtained from the first phase is used for building the initialization of the search process. The effectiveness of this method is demonstrated through empirical study on KDDCUP99 datasets [9].

II. THE KDDCUP99 DATASET

In the 1998 DARPA cyber attack detection evaluation program an environment [9] [10] was setup to acquire raw TCP/IP dump data for a network by simulating a typical U.S. Air Force LAN. The LAN was operated like a true environment, but being blasted with multiple attacks. For each TCP/IP connection, 41 various quantitative (continuous data type) and qualitative (discrete data type) features were extracted among the 41 features, 34 features are numeric and 7 features are symbolic. The data contains 24 attack types that could be classified into four main categories:

- DOS: Denial Of Service attack.
- R2L: Remote to Local (User) attack.
- U2R: User to Root attack.
- Probing: Surveillance and other probing.

*A. Denial of service Attack (DOS)*

Denial of service (DOS) is class of attack where an attacker makes a computing or memory resource too busy or too full to handle legitimate requests, thus denying legitimate user access to a machine.

*B. Remote to Local (User) Attacks*

A remote to local (R2L) attack is a class of attacks where an attacker sends packets to a machine over network, then exploits the machine's vulnerability to illegally gain local access to a machine.

*C. User to Root Attacks*

User to root (U2R) attacks is a class of attacks where an attacker starts with access to a normal user account on the system and is able to exploit vulnerability to gain root access to the system.

*D. Probing*

Probing is class of attacks where an attacker scans a network to gather information or find known vulnerabilities. An attacker with map of machine and services that are available on a network can use the information to notice for exploit.

TABLE I.   CLASS LABLE THAT APPEARS IN 10% DATA SET.

| Attack | No. of Samples | Class |
|---|---|---|
| Smurf | 280790 | DOS |
| Neptune | 107201 | DOS |
| Back | 2203 | DOS |
| Teardrop | 979 | DOS |
| Pod | 264 | DOS |
| Land | 21 | DOS |
| Normal | 97277 | Narmal |
| Satan | 1589 | Prob |
| Ipsweep | 1247 | Prob |
| Portsweep | 1040 | Prob |
| Nmap | 231 | Prob |
| Warezclient | 1020 | R2L |
| Guess_passwd | 53 | R2L |
| warezmaster | 20 | R2L |
| Imap | 12 | R2L |
| ftp_write | 8 | R2L |
| multihop | 7 | R2L |
| Phf | 4 | R2L |
| Spy | 2 | R2L |
| buffer_overflow | 30 | U2R |
| Rootkit | 10 | U2R |
| loadmodule | 9 | U2R |
| Perl | 3 | U2R |

This dataset contains 22 attack types which can be classified into five classes i.e. Normal, DOS Probe, U2R and R2L and sample size of each attack which is shown in table Table-I.

### III. INFORMATION GAIN

Let S be a set of training set samples with their corresponding labels. Suppose there are m classes and the training set contains $S_i$ samples of class *I* and S is the total number of samples in the training set. Expected information needed to classify a given sample is calculated by:

$$I(S_1, S_2, \ldots S_m) = -\sum S_i/S \ \log_2(S_i/S) \qquad (1)$$

A feature F with values $\{f_1, f_2, \ldots f_v\}$ can divided the training set into v subset $\{S_1, S_2, \ldots S_v\}$ where $S_j$ is the subset which has the value $f_j$ for feature F is

$$E(F) = \sum (S_{1j} + S_{2j} + \ldots + S_{mj})/S \ast I(S_1, S_2, \ldots S_m) \qquad (2)$$

Information gain for F be calculated
$$GAIN(F) = I(S_1, S_2, \ldots S_m) - E(F) \qquad (3)$$

Information gain [3] is calculated for class labels by employing a binary discrimination for each class. That is for each class, a dataset instance is considered in class, if it has a different label. Consequently, as opposed to calculating one information gain as general measure on the relevance of the feature for all classes, we calculate an information gain for each class. Thus, this signifies how well the feature can discriminate the given class (i.e. normal or an attack type) from other classes.

### IV. ENSEMBLE APPROACH

Filter [11][12][13] and wrapper models [14][15][16] and avoid the pre specification of a stopping criterion, the hybrid model [17][18] is recently proposed to handle large data sets. A typical hybrid algorithm makes use of both an independent measure and a mining algorithm to evaluate feature subsets: It uses the independent measure to decide the best subsets for a given cardinality and uses the mining algorithm to select the final best subset among the best subsets across different cardinalities.

Basically, it starts the search from a given subset $S_0$ (usually, an empty set in sequential forward selection) and iterates to find the best subsets at each increasing cardinality. In each round for a best subset with cardinality c, it searches through all possible subsets of cardinality c+1 by adding one feature from the remaining features. Each newly generated subset S with cardinality c+1 is evaluated by an independent measure M and compared with the previous best one. If S is better, it becomes the current best subset $s'_{best}$ at level c+1. At the end of each iteration a mining algorithm A is applied on $s_{best}$ at level c+1 and the quality of the mined result is compared with that from the best subset at level c. If $s_{best}$ is better, the algorithm continues to find the best subset at the next level.

#### A. Ensemble approach

In this approach, we combine the two algorithms for final feature selection. The filter phase select the features with highest information gain and guides the initialization of search process for wrapper phase whose output the final feature subset. The propose hybrid algorithm is given below:

Proposed ensemble Algorithm ( )

```
Input:Kdd[][];//kdd data set
Output:s_{best}[];//an optimal threshold

Number accuracy1,accuracy2,N=41,m=1,A,c1,
c2,Gain[];

1.   begin
2.   Fori=1 to N do
3.     Gain[i]=information[i];
4.   end;
5.   Sort Gain[i]in ascending order;
6.   s_{best}[1]=getFirstElement(gain[],FN);
7.   accuracy1 = KNN(s_{best}[]);
8.   forc=0to41begin
```

```
9.      A=getNextElement(Gain[],FN);
10.     s'_best[i] = s_best[];
11.     accuracy2 = KNN(s'_best[i]);
12.     if(accuracy2>accuracy1) then
            s_best[] = s'_best[];
        else return s_best[];
13.    end;
```

The above mentioned pseudo code is our proposed hybrid algorithm for feature selection. This algorithm makes use of both a filter algorithm and wrapper algorithm to evaluate feature subsets. It uses the filter algorithm to decide the best subsets for a given cardinality and uses wrapper algorithms to select the final best subset among the best subset across different cardinalities. While in our filter algorithm is used to rank the feature on the basis of information gain. A feature which is having higher information gain for the prediction of the class is ranked high among other features. Algorithm starts with one cardinality means starting subset contain only one feature having highest information gain. Wrapper algorithm is used to evaluate the subset on the basis of accuracy. The final feature subsets are passed through the K-nearest neighbor classifier for classification of attacks.

## V. K-NEAREST NEIGHBOR CLASSIFIER

The k-nearest neighbor (KNN) [19] classifier is amongst the simplest of all machine learning algorithms. An object is classified by a majority vote of its neighbors, with the object being assigned to the class most common amongst its *k* nearest neighbors. *k* is a positive integer, typically small. If *k* =1, then the object is simply assigned to the class of its nearest neighbor. In binary (two class) classification problems, it is helpful to choose *k* to be an odd number as this avoids tied votes. The same method can be used for regression, by simply assigning the property value for the object to be the average of the values of its *k* nearest neighbors. It can be useful to weight the contributions of the neighbors, so that the nearer neighbors contribute more to the average than the more distant ones.

The neighbors are taken from a set of objects for which the correct classification (or, in the case of regression, the value of the property) is known. This can be thought of as the training set for the algorithm, though no explicit training step is required. In order to identify neighbors, the objects are represented by position vectors in a multidimensional feature space. It is usual to use the Euclidean distance, though other distance measures, such as the Manhattan distance could in principle be used instead. The *k*-nearest neighbor algorithm is sensitive to the local structure of the data.

The performance of a KNN classifier is primarily determined by the choice of K as well as the distance metric applied. However, it has been shown in that when the points are no t uniformly distributed, predetermining the value of K becomes difficult. Generally, larger values of K are more immune to the noise presented, and make boundaries smooth between classes. As a result, choosing the same (optimal) K becomes almost impossible for different applications.

### A. training

The training phase of the algorithm consists only of storing the feature vectors and class labels of the training samples. In the actual classification phase, the test sample (whose class is not known) is represented as a vector in the feature space. Distances from the new vector to all stored vectors are computed and k closest samples are selected. There are a number of ways to classify the new vector to a particular class. One of the most used techniques is to predict the new vector to the most common class amongst the K nearest neighbors. A major drawback to using this technique to classify a new vector to a class is that the classes with the more frequent examples tend to dominate the prediction of the new vector, as they tend to come up in the K nearest neighbors when the neighbors are computed due to their large number. One of the ways to overcome this problem is to take into account the distance of each K nearest neighbors with the new vector that is to be classified and predict the class of the new vector based on these distance.

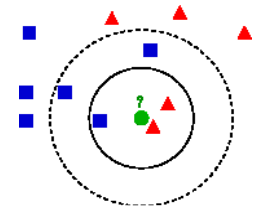

Figure 1: K-NN classification

In figure1 shown K-NN classification in which the test sample (green circle) should be classified either to first class of blue squares or the second class of red triangles. If k=3 it is classified to the second class because there are 2 triangle and only one square inside the inner circle. If k=5 it is classified to first class (3 squares vs. 2triangles inside the outer circle)

### B. Parameter selection

The best choice of k depends upon the data; generally, larger values of k reduce the effect of noise on the classification, but make boundaries between classes less distinct. A good k can be selected by various heuristic techniques, for example, cross-validation. The special case where the class is predicted to be the class of the closest training sample (when k=1) is called the nearest neighbor.

The accuracy of the k-NN algorithm can be severely degraded by the presence of noisy or irrelevant features, or if the feature scales are not consistent with their importance. Much research effort has been put into selecting or scaling features to improve classification. A particularly popular approach is the use of evolutionary algorithms to optimize feature scaling. Another popular approach is to scale features by the mutual information of the training data with the training classes.

## VI. EXPERIMENTS AND RESULTS

Information gain and KNN based hybrid approach is implemented in JAVA language. All experiments are run on a PC with a 1.73 GHz with 2 GB of RAM and running Windows XP. The KDDcup99 cyber attack detection bench consists of three components, which are shown in table II. In the International Knowledge Discovery and Data Mining Tools Competition only "10% KDD" dataset is employed for the purpose of training.

TABLE II. KDDCUP99 CYBER ATTACKDETECTION DATASET IN TERMS OF NUMBER OF SAMPLES.

| dataset | Normal | Probe | DOS | R2L | U2R |
|---|---|---|---|---|---|
| 10%KDD | 97277 | 4107 | 391458 | 1126 | 52 |
| Corrected KDD | 60593 | 4166 | 229853 | 16347 | 70 |
| Whole KDD | 972780 | 41102 | 3883370 | 1126 | 52 |

We applied the preprocessing algorithm to translate data in the format we need. In preprocessing steps we read data file line by line where each line contain a record. It replaces the comma by blanks and each feature is converted from text or symbolic from into numerical form. We place the data from file corrected.txt into two dimensional table KDD [][].

TABLE III. INFORMATION GAIN FOR ALL THE 41 FEATURS.

| S.No. | Name of features | Information Gain |
|---|---|---|
| 1 | Duration | 0.1437721 |
| 2 | Protocol type | 0.4980799 |
| 3 | Service | 0.975695 |
| 4 | Flag | 0.1563704 |
| 5 | Source bytes | 0.99074435 |
| 6 | Destination bytes | 0.722448 |
| 7 | Land | 1.38340243E-5 |
| 8 | Wrong fragment | 4.801972E-4 |
| 9 | Urgent | 1.114560E-4 |
| 10 | Hot | 0.007694 |
| 11 | Failed logins | 0.01446894 |
| 12 | Logged in | 0.99074435 |
| 13 | Compromised | 0.00281515 |
| 14 | Root shell | 0.00107513 |
| 15 | Su attempted | 3.1226067E-5 |
| 16 | Root | 8.1776882E-4 |
| 17 | File creations | 0.0011696 |
| 18 | Shell | 3.8327060E-4 |
| 19 | Access files | 0.00157854 |
| 20 | Outbound cmds | 0.0 |
| 21 | Is hot login | 0.99074435 |
| 22 | Is guest login | 0.9907443 |
| 23 | Count | 0.9907443 |
| 24 | Srv count | 0.9907443 |
| 25 | Serror rate | 0.0640963 |
| 26 | Srv serror rate | 0.0414653 |
| 27 | Rerror rate | 0.1222657 |
| 28 | Srv rerror rate | 0.1181657 |
| 29 | Same srv rate | 0.1570540 |
| 30 | Diff srv rate | 0.1571343 |
| 31 | Srv diff host rate | 0.1494001 |
| 32 | Dst host count | 0.2500145 |
| 33 | Dst host srv count | 0.9862856 |
| 34 | Dst host same srv rate | 0.3194438 |
| 35 | Dst host diff srv rate | 0.3868191 |
| 36 | Dst host same src port rate | 0.4627201 |
| 37 | Dst host srv diff | 0.2116295 |
| 38 | Dst host serror rate | 0.0808271 |
| 39 | Dst host srv serror rate | 0.0551282 |
| 40 | Dst host rerror rate | 0.1925867 |
| 41 | Dst host srv rerror rate | 0.1436175 |

We obtained feature gain from filter algorithm for all 41 features which is mentioned in table.III. We also store the records in the five files namely Normal.data, DOS.data, Probe.data, U2R.data and R2L.data.

TABLE IV. KDDCUP99 DATASET IN TERMS OF NUMBER OF SAMPLES.

| Dataset Size | Selected Features | Accuracy with 41 features | Accuracy with Selected features |
|---|---|---|---|
| 1000 | 5,12,21,22,23,24 | 72.21% | 80.09% |
| 10000 | 2,3,5,12,21,22,23,24 | 74.53% | 91.01% |
| 50000 | 2,3,6,5,12,21,22,23,24 | 73.78% | 91.7% |
| 100000 | 3,5,6,12,21,22,23,24 | 74.93% | 92.08% |
| 150000 | 2,3,4,5,6,12,21,22,23,24,31 | 75.09% | 92.09% |
| 250000 | 3,5,6,12,21,22,23,24,32,36 | 73.04% | 92.02% |

As KNN work first it takes a record from test dataset and calculates its distance from each records of training dataset. Then take 10 records of training dataset which have minimum distance. Then examine maximum of these 10 records belongs to which class. If the actual class of the test record is same as the resulting class then it is positive prediction. Same procedure run for each record in test dataset and at the end it finds out the accuracy by ratio of positive prediction and total number of records in the data set. If we have N records for test and after applying KNN we only fined x positive prediction, then accuracy is given by

$$Accuracy = x/N \quad Where\ x \leq N$$

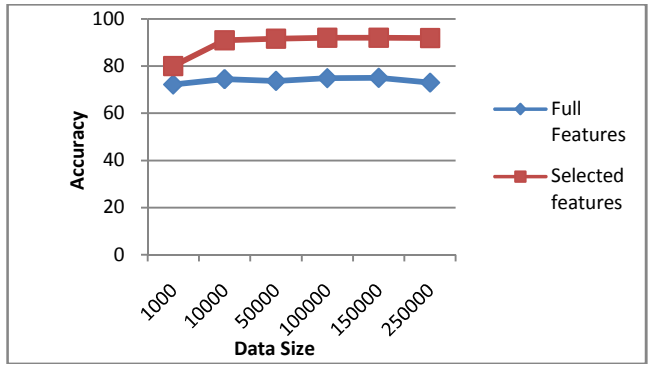

Figure 1. Comparision of Accuracy with full features and selected features by K-NN classifier.

The k-nearest neighbor is instance based classifier it randomly selects the different number of records from KDDCUP99 dataset to see the behavior of algorithm. When we apply the dataset of different sizes to the classifier and we compare the classifier accuracy for all 41 features and selected features as shown in Fig.1 accuracy of the selected features is increasing with the size of data. As shown in figure the accuracy of the classifier is increasing as with the data and it is higher than the accuracy of full features. We also analyze that all selected feature comes in the first 15 highest ranked features.

## VII. CONCLUSION

In this paper we use an ensemble approach for feature selection of KDDCUP99 dataset. Which consist of two techniques of feature selection that are filter and wrapper. The filter phase calculates the information gain of each feature and ranks them and gives input to wrapper phase. Wrapper phase searches for the final best subset of feature. The experiment demonstrate the effectiveness of our algorithm on KDDcup99 dataset which can be seen from the table IV the accuracy of the classifier with 41 features are less as compared to the accuracy of the same classifiers with selected features. As shown in the table IV there is no significant improvement in the detection accuracy (when 1000 records, accuracy improved from 72.21% to 80.09%) when attack records are small. But when the data size of attacks are huge (say 250000 records accuracy improved from 73.04% to 92.02%) there is an improvements in detection accuracy. Although we are getting improved cyber attack detection accuracy there is still a need for improvement the detection accuracy. We will investigate new technique for further improvement.

## APPENDIX

APPENDIX-1 CLASS LABLE THAT APPEARS IN 10% DATA SET.

| S.No | Name of features | Description | Data type |
|---|---|---|---|
| 1 | Duration | Duration of the connection | C |
| 2 | Protocol type | Connection protocol | D |
| 3 | Service | Destination service | D |
| 4 | Flag | Status flag of the connection | D |
| 5 | Source bytes | Bytes sent from source to destination | C |
| 6 | Destination bytes | Bytes sent from destination to source | C |
| 7 | Land | 1 if connection is from/to the same host/port; 0 otherwise | D |
| 8 | Wrong fragment | Number of wrong fragment | C |
| 9 | Urgent | Number of urgent packets | C |
| 10 | Hot | Number of "hot" indication | C |
| 11 | Failed logins | Number of failed logins | C |
| 12 | Logged in | 1 if successfully logged in;0 otherwise | D |
| 13 | # compromised | Number of "compromised" conditions | C |
| 14 | Root shell | 1 if root shell is obtained;0 otherwise | C |
| 15 | Su attempted | 1 if "su root" command attempted 0 otherwise | C |
| 16 | Root | No. of "root" accesses | C |
| 17 | File creations | No. of file creation operation | C |
| 18 | Shell | Number of shell prompt | C |
| 19 | Access files | Number of operations on access control files | C |
| 20 | Outbound cmds | Number of outbound commands in an ftp session | C |
| 21 | Is hot login | 1 if the login belongs to the "hot" list; 0 otherwise | D |
| 22 | Is guest login | 1 if the login is a guest login 0 otherwise | D |
| 23 | Count | Number of connections to the same host as the current connection in the past 2 seconds | C |
| 24 | Srv count | Number of connection to the same service as the current connection in past 2 seconds | C |
| 25 | Serror rate | % of connection that have "SYN" error | C |
| 26 | Srv serror rate | % of connection that have "SYN" error | C |
| 27 | Rerror rate | % of connection that have "REJ" error | C |
| 28 | Srv rerror rate | % of connection that have "REJ" error | C |
| 29 | Same srv rate | % of connection to the same service | C |
| 30 | Diff srv rate | % of connection to different service | C |
| 31 | Srv diff host rate | % of connection to different host | C |
| 32 | Dst host count | Count of connection having same dest hot | C |
| 33 | Dst host srv count | Count of connection having the same destination host and using same service | C |
| 34 | Dst host same srv rate | % of connection having the same destination host and using same service | C |
| 35 | Dst host diff srv rate | % of different service on the current host | C |
| 36 | Dst host same src port rate | % of connection to the current hot having same src port | C |
| 37 | Dst host srv diff | % of connection to the same service coming from different host | C |
| 38 | Dst host serror rate | % of connection to the current host that have an S0 error | C |
| 39 | Dst host srv serror rate | % of connection to the current host and specified service that have an S0 error | C |
| 40 | Dst host rerror rate | % of connection to the current host that have an RST error | C |
| 41 | Dst host srv rerror rate | % of connection to the current host and specified service that have an RSTerror | C |

C- Continuous, D-Discrete

## AUTHORS PROFILE

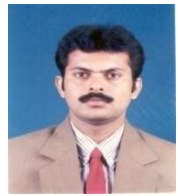

**Shailendra Singh** Asistant Professor in Department of Information Technology at Rajiv Gandhi Technological University, Bhopal, India. He has publised more than 14 papers in international journals and conference proceedings His research interest include datamining and network security.He is a life member of ISTE, Associte member of Institution of Engineers (India) and member of International Association of Computer Science and Information Technology (IACSIT) Singapore.

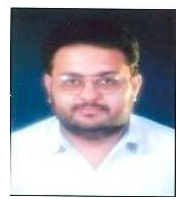

**Dr. Sanjay Silakari** Professor and Head Department of Computer Science and Engineering at Rajiv Gandhi Technological University, Bhopal, India. He has awarded Ph.D. degree in Computer Science & Engg. He posses more than 16 years of experience in teaching under-graduate and post-graduate classes. He has publised more than 60 papers in international, national journals and conference proceedings.